\documentclass[12pt]{article}
\usepackage{graphicx}
\usepackage{amssymb}
\newlength{\extraspace}
\newlength{\extraspaces}
\newcommand{\be}{\begin{equation}
\addtolength{\abovedisplayskip}{\extraspaces}
\addtolength{\belowdisplayskip}{\extraspaces}
\addtolength{\abovedisplayshortskip}{\extraspace}
\addtolength{\belowdisplayshortskip}{\extraspace}}
\newcommand{\ee}{\end{equation}}

\newcommand{\ba}{\begin{eqnarray}
\addtolength{\abovedisplayskip}{\extraspaces}
\addtolength{\belowdisplayskip}{\extraspaces}
\addtolength{\abovedisplayshortskip}{\extraspace}
\addtolength{\belowdisplayshortskip}{\extraspace}}
\newcommand{\ea}{\end{eqnarray}}

\newcommand{\newsection}[1]{
\vspace{12mm}
\pagebreak[3]
\addtocounter{section}{1}
\setcounter{equation}{0}
\setcounter{subsection}{0}
\setcounter{subsubsection}{0}
\noindent{\bf \thesection. #1}
\nopagebreak
\medskip
\nopagebreak
\pdfbookmark[1]{#1}{#1}}

\newcommand{\newsubsection}[1]{
\vspace{0.8cm}
\pagebreak[3]
\addtocounter{subsection}{1}
\noindent{\it \thesubsection. #1}
\nopagebreak
\vspace{2mm}
\nopagebreak
}

\newcounter{saveeqn}

\newcommand{\dif}{\mathrm{d}}

\newcommand{\Eqref}[1]{(\ref{#1})}
\newcommand{\half}{\frac{1}{2}}
\newcommand{\expo}[1]{\mathrm{e}^{#1}}
\newcommand{\brac}[1]{\left(#1 \right)}
\newcommand{\sbrac}[1]{\left[#1\right]}

\newcommand{\hm}{h_\mathrm{m}^2}
\newcommand{\bm}{b_\mathrm{m}^2}
\newcommand{\ttau}{\frac{\theta}{\rho_0}}
\newcommand{\hub}{\frac{a'}{a}}
\newcommand{\eq}{\;=\;}

\usepackage{amsmath}
\usepackage{tabularx}
\usepackage{multirow}
\usepackage{amssymb}
\usepackage{psfrag}
\usepackage{setspace}
\usepackage[colorlinks=true,bookmarks=true]{hyperref}

\begin{document}
\addtolength{\baselineskip}{1.5mm}

\thispagestyle{empty}
\begin{flushright}

\end{flushright}
\vbox{}
\vspace{2cm}

\begin{center}
{\LARGE{Warped branches of flux compactifications
        }}\\[16mm]
{Yen-Kheng Lim\footnote{Email address: phylyk@nus.edu.sg}}
\\[6mm]
{\it Department of Physics,
National University of Singapore, 
Singapore 119260}\\[15mm]

\end{center}
\vspace{2cm}

\centerline{\bf Abstract}
\bigskip
\noindent
We consider Freund-Rubin-type compactifications which are described by $(p+q)$-dimensional Einstein gravity with a positive cosmological constant and a $q$-form flux.  Using perturbative expansions of Kinoshita's ansatz for warped $\mathrm{dS}_p\times S^q$ and $\mathrm{AdS}_p\times S^q$ spacetimes, we obtain analytical solutions describing the warped branches and their respective phase spaces.  These equations are given by inhomogeneous Gegenbauer differential equations which can be solved by the Green's function method.  The requirement that the Green's functions are regular provides constraints which determine the structure of the phase space of the warped branches.  We apply the perturbation results to calculate the thermodynamic variables for the warped $\mathrm{dS}_p\times S^q$ branch.  In particular, the first law of thermodynamics can be reproduced using this method.
\newpage

\newsection{Introduction}

Compactified solutions for de Sitter (dS) and anti-de Sitter (AdS) spacetimes have been the subject of interest from various viewpoints.  For instance, $(p+q)$-dimensional  spacetimes of product form $\mathrm{AdS}_p\times\mathcal{M}_q$ are frequently considered in the context of supergravity and string/M-theory, as well as being an important system for the AdS/CFT correspondence.  On the other hand, as scenarios of inflationary cosmology and the present day acceleration are approximately modeled by the de Sitter spacetime, then $\mathrm{dS}_p\times\mathcal{M}_q$ comes under important consideration when one attempts to incorporate cosmological models into higher-dimensional theories of (quantum) gravity.  Thus it is of much interest to understand the physics required and caused by such scenarios.
\par
One of the important questions to address is the stability of such systems.  For instance, the question of stability is important for cosmology as it is well-known that the effects of cosmological perturbations from matter fields play a significant role in the early universe.  In the context of higher-dimensional supergravity, Freund and Rubin \cite{Freund:1980xh} have shown that the existence of a higher-dimensional flux will induce a natural compactification of the extra dimensions.  If the spacetime is maximally symmetric, the compactified solution is of the form $\mathrm{AdS}_p\times S^q$, where $p$ and $q$ are respectively the number of macroscopic and internal dimensions.  This was explored further by De Wolfe et al.~\cite{DeWolfe:2001nz} who studied the stability of general $\mathrm{AdS}_p\times\mathcal{M}_q$ configurations, where $\mathcal{M}_q$ is an Einstein space with positive curvature.  In these theories, the negative curvature of the external spacetime is due to the presence of a $q$-form flux in the internal space.
\par
If a positive bulk cosmological constant is included in the theory, more interesting cases can arise.  For a sufficiently small flux strength, the curvature of the external manifold can be positive and hence describes the de Sitter spacetime.  Since present cosmological observations and early inflationary epochs are modeled by the de Sitter spacetime, this allows us to consider flux compactifications within cosmological models.  For example, such a model has been considered by Carroll et al.~\cite{Carroll:2009dn} to construct a mechanism where the $\mathrm{dS}_p\times S^q$ geometries can dynamically arise from a spontaneous compactification of an initially uncompactified $(p+q)$-dimensional background.
\par
As mentioned above, a problem frequently considered in cosmology is that of perturbations in cosmological models, which are primarily studied to understand the formation of large scale structure in the universe and also to investigate the stability of compactifications in cosmological solutions.  This was considered in the context of Freund-Rubin-type compactifications by Martin \cite{Martin:2004wp}, which is related to earlier work by Bousso et al.~\cite{Bousso:2002fi}, Contaldi et al.~\cite{Contaldi:2004hr} and Frolov \cite{Frolov:2003yi}.  Martin's result is that for small flux, configurations in the form $\mathrm{dS}_p\times S^q$ are unstable against homogeneous perturbations, and beginning in the case $q=4$, for large flux, higher modes of perturbations ($l=2,3,\ldots$) begin to enter.  For the case $q\geq 5$, there are no more stable modes for $\mathrm{dS}_p\times S^q$.
\par
The onset of the threshold unstable mode may signal a presence of a new branch of solutions which are possibly non-uniform.  In Ref.~\cite{Kinoshita:2007uk}, Kinoshita looked for one of these solutions by generalizing $\mathrm{dS}_p\times S^q$ into an ansatz corresponding to a warped product of the form
\begin{align}
  \dif s^2 = A(y)^2g_{\mu\nu}(x)\dif x^\mu\dif x^\nu + g_{mn}(y)\dif y^m\dif y^n\,, 
\end{align}
where $x^\mu$ denote coordinates of the external spacetime and $y^m$ denote coordinates in the compact space.  Using this ansatz, a new branch was found numerically for $q=4$ where the compactified space is a deformed sphere.  More recently, Minamitsuji and Uzawa \cite{Minamitsuji:2011gp,Minamitsuji:2011xs} obtained warped de Sitter compactifications with the topology $\mathrm{dS}_p\times S^1\times S^{q-1}$.  In contrast to the one studied by Kinoshita, exact solutions were obtained for this case.
\par
Interestingly, the results reviewed above parallels the recent developments regarding the stability of higher-dimensional black holes.  In 2003, Emparan and Myers \cite{Emparan:2003sy} showed that ultra-spinning Myers-Perry black holes are subject to an instability, which led to a proposed construction of a black hole phase diagram whereby a new branch of ``lumpy'' black holes are postulated to emanate from the Myers-Perry branch \cite{Emparan:2007wm,Emparan:2010sx}.
\par
Subsequently, numerical analysis on various cases of Myers-Perry black holes have identified the points of marginal stability (which are points where the new branches are expected to appear) in the respective phase diagrams of singly-spinning black holes \cite{Dias:2009iu,Dias:2010maa}, and black holes with equal angular momenta \cite{Dias:2010eu}.  Due to the extreme complexity of the Einstein equations for such systems, most of the results were obtained using advanced numerics and heuristic constructions.
\par
The Freund-Rubin equations we are considering here have a slight advantage over black holes due to the fact that the instability appears in the scalar sector with respect to $\mathrm{dS}_p\times S^q$, thus making it more tractable to analysis.  The introduction of warping in Ref.~\cite{Kinoshita:2007uk} singles out only one direction where $S^q$ symmetry is broken, while preserving the others.  Such a system was solved numerically by Kinoshita to demonstrate the existence of a new branch of warped solutions for $p=4$, $q=4$, and a phase diagram is obtained.\footnote{The Freund-Rubin phase diagrams considered here will be parametrized by the Hubble constant and flux strength.}  Subsequent work by Kinoshita and Mukhoyama \cite{Kinoshita:2009hh} shows that the entropy of the warped branch is higher than the unwarped branch whenever the latter suffers from instability, providing further support to the \emph{correlated stability conjecture} by Gubser and Mitra \cite{Gubser:2000mm} which was originally introduced for black branes with infinite translational symmetry.  
\par
In this paper, using perturbative methods, we obtain analytical expressions to describe the phase diagrams for the warped branches.  As we will show below, in the phase space, we are able to write down the equation for the warped branch as a series expansion about the point of marginal stability.  The result we obtained supports the numerical analysis of \cite{Kinoshita:2007uk}, particularly, for $q=4$.  Continuing the analysis to $q=5$, $6$ and $7$ we calculate the full phase diagrams numerically and also find agreement with the perturbative analysis.  This procedure is analogous to the work by Gubser \cite{Gubser:2002} in the study of non-uniform black branes.  However for black branes even the perturbative expansions require numerical solutions.  In our case, the equations are sufficiently simple and analytical solutions can be obtained.  By analytical continuation of the Hubble parameter, the perturbative analysis can be easily extended to include anti-de Sitter spacetimes.  Thus both $\mathrm{dS}_p\times S^q$ and $\mathrm{AdS}_p\times S^q$ solutions can be treated equally well.\footnote{It is important to note that Kinoshita's ansatz corresponds to static solutions, thus the perturbative solutions only indicates the existence of a static threshold mode.  For the de Sitter solutions, this coincides with the onset of dynamical instability.  However for the anti-de Sitter case, in order to have a physical dynamical instability, the perturbative modes must also violate the well-known Breitenlohner-Freedman bound; such considerations are beyond the scope of this paper.}
\par
This paper is organized as follows: in Section 2 we review the theory for warped compactifications with flux, as well as setting up and defining the notations to be used throughout the rest of the paper.  In the same section, we will also reiterate the main results of Refs. \cite{Bousso:2002fi,Martin:2004wp} regarding the stability of $\mathrm{dS}_p\times\mathcal{M}_q$.  In Section 3, we present the perturbative analysis of Kinoshita's ansatz.  In particular, we will focus on the even modes ($l=2N$ where $N$ is an integer) of the perturbative solutions.  We show that at first order, they are equivalent to the results in \cite{Bousso:2002fi,Martin:2004wp}.  Proceeding to second and higher orders we obtain the analytical structure of the phase diagram near the points where the warped branch emanates from the unwarped branch.  There is a subtlety in the procedure for odd $l$ cases, which will be addressed in Section 4.  Subsequently in Section 5 we apply our results to calculate thermodynamic quantities for $\mathrm{dS}_p\times S^q$ and reproduce the first law of de Sitter thermodynamics.  We conclude the paper in Section 6 and present numerical results for the warped branches in $q=5$, $6$ and $7$ in  Appendix A.  In Appendix B we describe the procedure to derive the perturbation equation used in Section 3.

\newsection{Freund-Rubin-type compactifications}

\newsubsection{The action}
\pdfbookmark[2]{The action}{The action}

We begin by reviewing the theory of Freund-Rubin-type compactifications.  They describe higher dimensional theories where the extra compact dimensions are stabilized by the presence of a $q$-form flux.  We will also include the presence of a bulk cosmological constant $\Lambda$.  It was first observed by Freund and Rubin \cite{Freund:1980xh} that the presence of $F_{(q)}$ can induce a natural compactification such that the solutions can be written as a product $\mathcal{M}_p\times\mathcal{N}_q$, where $\mathcal{M}_p$ is the external $p$-dimensional spacetime with Lorentzian signature and $\mathcal{N}_q$ is the compact $q$-dimensional space with Euclidean signature.\par
The starting point for the theory is the following action (in units where $G=c=1$)
\begin{align}
 I=\frac{1}{16\pi}\int\dif^px\,\dif^qy\;\sqrt{-g}\brac{R-2\Lambda-\frac{1}{q!}F^2_{(q)}}\,, \label{action}
\end{align}
where $R$ is the scalar curvature, $F_{(q)}$ is the $q$-form field, and $\Lambda$ is the positive cosmological constant.  We will use lowercase Greek indices $(\mu,\nu,\ldots)$ to denote coordinates in the $p$-dimensional external spacetime, and lowercase Latin indices $(m,n,\ldots)$ to denote coordinates in the $q$-dimensional compact space.  Uppercase Latin indices $(M,N,\ldots)$ will denote all coordinates from the entire $(p+q)$ dimensions.\par
Varying the action \Eqref{action} gives the Einstein-Maxwell equations
\begin{align}
 &R_{MN} = \frac{1}{(q-1)!}F_{ML_2\ldots L_{q}}{F_N}^{L_2\ldots L_{q}} - \frac{q-1}{p+q-2}\frac{1}{q!}F^2g_{MN}+\frac{2}{p+q-2}\Lambda g_{MN}\,, \nonumber\\
 &\nabla_M F^{ML_2\cdots L_{q}} =0\,. \label{EOM}
\end{align}
In this work we are interested in solutions where the external spacetime is $p$-dimensional (anti)-de Sitter space, with a $q$-dimensional internal space.  Following Kinoshita we consider the case where the internal space has $SO(q)$ isometry.  Such configurations can be described by Kinoshita's ansatz \cite{Kinoshita:2007uk}
\begin{align}
 \dif s^2=\;&\expo{2\Phi(\theta)}\sbrac{-\brac{1-h^2r^2}\dif t^2+\brac{1-h^2r^2}^{-1}\dif r^2+r^2\dif\Omega^2_{(p-2)}}\nonumber\\
          &+\expo{-\frac{2p}{q-2}\Phi(\theta)}\sbrac{\dif\theta^2+a(\theta)^2\dif\Omega^2_{(q-1)}}\,, \label{kinometric}
\end{align}
where $h$ is the Hubble parameter and $\dif\Omega^2_{(k)}$ denotes the metric of a $k$-dimensional unit sphere.  The $q$-form field that solves the Maxwell equations can be parametrized by its strength $b$ and is given by $F_{(q)}=ba^{q-1}\expo{-\frac{2p(q-1)}{q-2}\Phi}\dif\theta\wedge\dif\Omega_{(q-1)}$.  The warp factor $\Phi(\theta)$ and $a(\theta)$ are functions that depend only on one (finite) internal coordinate $\theta$, which lies in the interval $\theta_-\leq\theta\leq\theta_+$.  Loosely speaking, we may call $\theta_-$ and $\theta_+$ as the `south' and `north' poles respectively.  We further let $\theta_*$ be the position of the `equator', whose definition will be made more precise below.  In this case, the Maxwell equations are trivially satisfied, while the Einstein equations reduce to
\begin{align}
 \Phi''+(q-1)\hub\Phi'&=(p-1)h^2\expo{-\frac{2(p+q-2)}{q-2}\Phi}+\frac{q-1}{p+q-2}b^2\expo{-\frac{2p(q-1)}{q-2}\Phi}-\frac{2\Lambda}{p+q-2}\expo{-\frac{2p}{q-2}\Phi}\,, \label{eom1} \\
 a''+\frac{1-a'^2}{a}&=-\frac{p(p+q-2)}{(q-2)^2}a\Phi'^2\,, \label{eom2}
\end{align}
with a constraint equation
\begin{align}
  (q-1)(q-2)\brac{\frac{a'^2-1}{a^2}}=\;&\frac{p(p+q-2)}{(q-2)}\Phi'^2+p(p-1)h^2\expo{-\frac{2(p+q-2)}{q-2}\Phi}\nonumber\\
                                 &+b^2\expo{-\frac{2p(q-1)}{q-2}\Phi}-2\Lambda\expo{-\frac{2p}{q-2}\Phi}\,, \label{constraint}
\end{align}
where primes denote derivatives with respect to $\theta$.  The cosmological constant can be normalized to unity by rescaling $b\rightarrow b\Lambda^{1/2}$, $ h\rightarrow h\Lambda^{1/2}$, $a\rightarrow a\Lambda^{-1/2}$ and $ \theta\rightarrow\theta\Lambda^{-1/2}$; this rescaling implies that all length scales are measured in units of (anti)-de Sitter curvature.
\par
In the metric \Eqref{kinometric}, there is still some freedom which is yet to be fixed, namely, the following transformations parametrized by a constant $\Phi_0$
\begin{align}
  \Phi\rightarrow\Phi+\Phi_0\,,\hspace{0.2cm}a&\rightarrow a\expo{\frac{p}{q-2}\Phi_0}\,,\hspace{0.2cm}h\rightarrow h\expo{\Phi_0}\,,\hspace{0.2cm}b\rightarrow b\expo{p\Phi_0}\,, \nonumber\\
  \theta&\rightarrow\theta\expo{\frac{p}{q-2}\Phi_0}\,,\hspace{0.2cm}t\rightarrow t\expo{-\Phi_0}\,, \label{invariance}
\end{align}
which leaves the metric invariant. In practice, this translates into a freedom in fixing the boundary condition for $\Phi$ when integrating \Eqref{eom1}.  Obviously this has no physical significance as the transformation in \Eqref{invariance} leaves the metric and the physical observables unchanged.  However, when performing numerical calculations, this freedom has to be exploited correctly to ensure that we discover all branches of solutions.  In particular, a suitable choice of $\Phi_0$ is important so that the numerical analysis is able to discover the odd $l$ branches (see Appendix A).

\newsubsection{The FR branch and its stability}
\pdfbookmark[2]{The FR branch and its stability}{The FR branch and its stability}

The Einstein-Maxwell equations \Eqref{eom1} and \Eqref{eom2} admit a trivial solution of the form $\mathrm{(A)dS}_p\times S^q$, which we will refer to as the Freund-Rubin (FR) branch.  In this case, the solutions are
\begin{align}
 \Phi = 0\,,\hspace{0.5cm} a(\theta)=\rho_0\cos{\frac{\theta}{\rho_0}}\,. \label{FRsoln}
\end{align}
Here, the north and south poles are $\theta_\pm=\pm\rho_0\pi/2$, and the equator is $\theta_*=0$. The components of the $q$-form field reduce to
\begin{align}
F_{L_1\ldots L_q} &= \left\{
		  \begin{array}{cl}
		   b\epsilon_{l_1\ldots l_q} & \mbox{if } L_i=l_i\,,\label{FR2} \\
                   0 & \mbox{otherwise.}
		  \end{array}
                \right.
\end{align}
This solution is parametrized by three quantities: the Hubble parameter $h^2$, the flux strength $b^2$, and compactification radius $\rho_0$.  We further note that the external space is de Sitter if $h^2>0$, Minkowski if $h^2=0$, or anti-de Sitter if $h^2<0$.  
The parameters $\{b^2,h^2,\rho_0\}$ are constrained by the Einstein-Maxwell equations.  To see this we simply substitute the trivial solution \Eqref{FRsoln} into Eqs.~\Eqref{eom1} and \Eqref{eom2}.  This results in the following algebraic relations
\begin{align}
 (p-1)(p+q-2)h^2 + (q-1)b^2 &= 2\Lambda\,, \label{FRbranch1}\\
 (q-1)(p+q-2)\rho_0^{-2} - (p-1)b^2 &= 2\Lambda\,, \label{FRbranch2}
\end{align}
leaving one free parameter to describe the Freund-Rubin solutions.  We can therefore describe this family of FR solutions as curves in the $(b^2,h^2)$ phase space, which, according to the linear relation \Eqref{FRbranch1}, are simply straight lines.
\par
The FR branch contains two channels of instabilities \cite{Bousso:2002fi,Martin:2004wp}; the first is due to the homogeneous mode ($l=0$), which occurs for
\begin{align}
 b^2\leq\frac{2\Lambda}{(p-1)(q-1)}\,. \label{homoinstab}
\end{align}
This instability is due to homogeneous perturbation of the compact sphere, which results in a uniform expansion or contraction of the compact space.  Therefore, this instability does not change the shape of the compactification.  \par
The second instability, which is the main concern of this paper, are the inhomogeneous perturbations ($l=2,3,\ldots$) occuring when \cite{Martin:2004wp,Bousso:2002fi}
\begin{align}
 b^2\geq \frac{2\Lambda[l(l+q-1)-2q+2]}{(p-1)\sbrac{2(q-1)^2-l(l+q-1)}}\,, \label{instab}
\end{align}
where $-l(l+q-1)$ are the eigenvalues of the Laplace operator on $S^q$, corresponding to the eigenmode $l$.  If the flux satisfies \Eqref{instab}, the linearized analysis show that there will be modes that deform the compact sphere which grow exponentially in time.  The threshold stationary mode is the limiting case where equality occurs in \Eqref{instab}.  This signals the presence of a new branch of solutions with deformed compact spheres according to the shape of its mode $l$.  These points are called \emph{marginal stability points}, and will be denoted by $(b^2_\mathrm{m},h^2_\mathrm{m})$. \par
Thus, to obtain stable FR solutions, one must choose the parameters $(b^2,h^2)$ and dimensions $p$ and $q$ to avoid the inequalities in Eqs.~\Eqref{homoinstab} and \Eqref{instab}.  For dimension $q\geq5$, the inequalities \Eqref{homoinstab} and \Eqref{instab} overlap for all $h^2$ and $b^2$, thus there are no more stable FR solutions for $q\geq5$.\par
From here onwards, we consider the inhomogeneous instabilities and the resulting new branches.  The $(b^2,h^2)$ phase space gives us an intuitive interpretation of the stability of the solutions: when $b^2$ exceeds a particular threshold point indicated by Eq.~\Eqref{instab}, the FR solution becomes unstable to inhomogeneous perturbations.  This leads to the possibility of a new class of stable solutions which are distinct from the FR branch.  The intersection between the FR class and the new class should occur at the point of marginal stability $(b^2_\mathrm{m},h^2_\mathrm{m})$, where instability begins to arise.  Table \ref{points}, based on the calculations of \cite{Martin:2004wp}, shows various points of marginal instability for different cases of $q$.  In particular, we can easily see that with the increasing number of compact dimensions $q$, the number of unstable modes increases as well.\footnote{The points with negative values of $b^2$ are excluded as being unphysical.}

\begin{table}
\begin{center}
\renewcommand{\arraystretch}{1.5}
 \begin{tabular}{ccccc}
	\hline
             & $q=4$                               & $q=5$                               & $q=6$  & $q=7$  \\ \hline
     $l=2$:  & $\brac{\frac{1}{3},\frac{1}{18}}$ & $\brac{\frac{2}{15},\frac{22}{315}}$ & $\brac{\frac{2}{27},\frac{11}{162}}$& $\brac{\frac{1}{21},\frac{4}{63}}$   \\
     $l=3$:  & $(\infty,-\infty)$ & $\brac{\frac{26}{33},-\frac{38}{693}}$ & $\brac{\frac{14}{39},\frac{1}{117}}$& $\brac{\frac{2}{9},\frac{2}{81}}$    \\
     $l=4$:  & $\brac{-\frac{22}{15},\frac{16}{45}}$ & $(\infty,-\infty)$ & $\brac{\frac{26}{21},-\frac{11}{63}}$& $\brac{\frac{7}{12},-\frac{1}{18}}$   \\
     $l=5$:  & $\brac{-\frac{34}{33},\frac{28}{99}}$ & $\brac{-\frac{74}{39},\frac{374}{819}}$ & $(\infty,-\infty)$ & $\brac{\frac{86}{51},-\frac{46}{153}}$ \\ \hline
 \end{tabular}
 \caption{The points of marginal instability, $\brac{b^2_{\mathrm{m}}/\Lambda,h^2_{\mathrm{m}}/\Lambda}$  for $p=4$ and various $l$ and $q$}
 \label{points}
\end{center}

\end{table}

\newsubsection{The warped branch}
\pdfbookmark[2]{The warped branch}{The warped branch}

For the case $q=4$, Kinoshita has solved Eqs.~\Eqref{eom1} and \Eqref{eom2} numerically and found that the warped branch is also a single-parameter family which intersects the FR branch at the point of $l=2$ marginal instability.  For convenience, we reproduce the phase diagram in Figure \ref{q4branches}.  We also produce phase diagrams for higher $q$ cases in Appendix A for $q=5$, $6$ and $7$ (Figure \ref{q567-branches}).  The boundary conditions required to solve Eq.~\Eqref{eom1} and \Eqref{eom2} are:
\ba
  a(\theta_\pm)\eq0\,,\hspace{0.5cm}\left|a'(\theta_\pm)\right|=1\,,\hspace{0.5cm}\Phi'(\theta_\pm)=0\,. \label{BC}
\ea
The zero of the function $\Phi$ can be arbitrarily chosen due to freedom given by Eq.~\Eqref{invariance}.  In obtaining the solutions in Figure \ref{q4branches}, Kinoshita made the following choice \cite{Kinoshita:2007uk}:
\begin{align}
  \Phi(\theta_*)\eq0\,. \label{scheme1}
\end{align}
For the case of warped branches, we define the equator $\theta=\theta_*$ as the location where $a'(\theta_*)=0$.  Correspondingly $a(\theta_*)\equiv\rho$ will be called the `compactification radius', in analogy to the FR branch.
\begin{figure}
	\begin{center}
		\includegraphics[angle=-90,scale=0.5]{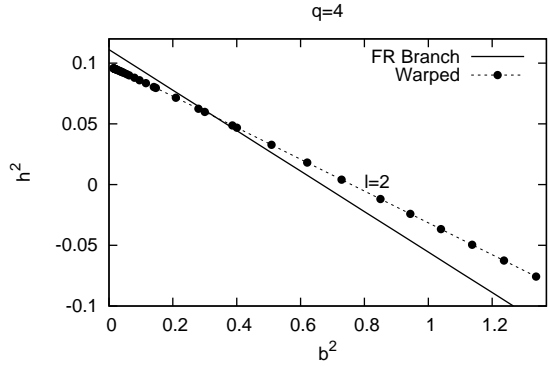}
		\caption{$b^2$ vs $h^2$ phase diagram of solutions $q=4$, first obtained by Kinoshita in Ref.~\cite{Kinoshita:2007uk}}
		\label{q4branches}	
	\end{center}
\end{figure}

\newsection{Perturbative expansion of the Kinoshita ansatz: even $l$ cases}
 
The warped branches obtained by Kinoshita in Figure \ref{q4branches}, and the higher $l$ modes shown in Figure \ref{q567-branches} appear to be straight; thus it was conjectured in \cite{Kinoshita:2007uk} that this branch may also be linear in $h^2$ and $b^2$, similar to Eq.~\Eqref{FRbranch1}.  However, as we will see below, this is not the case as this branch is actually slightly curved.  We will show this by obtaining an approximate analytical expression for the warped branch and show that there is a nonzero quadratic dependence on $\delta b^2$, where $\delta b^2$ is the amount of deviation away from the FR branch (this will be defined more precisely below).  We first focus on the cases where $l$ is even, i.e., $l=2N$ where $N$ is a positive integer.  The perturbations in the odd modes ($l=2N+1$) introduce a certain subtlety which we will address in Section 4.  We will also derive analytical solutions for $\Phi$ and $a$ under the approximation where $\Phi$ is small.  This method reproduces the points of marginal stability initially obtained by Bousso et al.~\cite{Bousso:2002fi} and Martin \cite{Martin:2004wp}, therefore we shall refrain from assuming their results.  \par
In this treatment, the warped branches in $(b^2,h^2)$ phase space can be written as an expansion from a (yet unspecified) point $(b^2_0,h^2_0)$ on the FR branch using $\delta b^2$ as the expansion parameter.  Therefore we write
\begin{align}
 b^2=b^2_0+\delta b^2\,,\hspace{0.5cm} h^2=h^2_0+w_1\delta b^2+w_2\brac{\delta b^2}^2+\ldots \label{delta_b2}
\end{align}
where the quantities $b_0^2$, $h_0^2$, $w_1$, $w_2\ldots$ will be determined from the perturbed equations.
Before proceeding to the details, we briefly outline the procedure that will be used to calculate $w_1$ and $w_2$:

\begin{itemize}
 \item Firstly, we write $\Phi$ and $a$ as a perturbation series expanded about the FR branch \Eqref{FRsoln}.
 \item In the first-order perturbation of \Eqref{eom1}, the requirement that the solutions are regular determines the points of marginal stability $(b^2_\mathrm{m},h^2_\mathrm{m})$.
 \item The second-order perturbation of \Eqref{eom1} results in an inhomogeneous differential equations whose source term is parametrized by $w_1$.  Demanding the solutions to be regular constrains $w_1$ to a particular value, giving the first order coefficient to the phase space equation \Eqref{delta_b2}.
 \item The same step applied to the third-order equations will correspondingly give the value of $w_2$.
\end{itemize}
Our starting point is to expand Eqs.~\Eqref{eom1} and \Eqref{eom2} about the FR solution \Eqref{FRsoln}.  In the warped solutions, the compact space is no longer a perfect sphere, and therefore the compactification radius is no longer a constant.  However, recalling the previous section, we define a quantity called `compactification radius' as $a(\theta_*)\equiv \rho$, and thus the perturbation solutions will introduce higher order corrections in the form $\rho=\rho_0+\rho_1\delta b^2+\rho_2\brac{\delta b^2}^2+\ldots$
\par
The problem we are considering in perturbation theory is to investigate the behavior of $\Phi$ and $a$ as the parameters $\brac{b^2,h^2}$ move away from the FR branch in accordance to Eq.~\Eqref{delta_b2}.  We begin by writing $\Phi$ and $a$ as a perturbation series under the parameter $\delta b^2$ about the FR solution:
\begin{align}
 \Phi=\;&0+\Phi_1\delta b^2+\Phi_2\brac{\delta b^2}^2+\Phi_3\brac{\delta b^2}^3+\ldots \label{phiseries} \\
 a=\;&\rho_0\cos\ttau+a_1\delta b^2 + a_2\brac{\delta b^2}^2+\ldots   \label{alphaseries}\,
\end{align}
Some care must be taken when choosing the appropriate coordinate system to use in the perturbation analysis.  It is important to note that the coordinate $\theta$ that appears in Eq.~\Eqref{alphaseries} has dimension of length, and hence will be physically deformed when the FR branch is perturbed.  In particular, we may expect the locations of the `poles' to be shifted by the perturbations.  This introduces an ambiguity in terms of the coordinate locations where the boundary conditions are to be imposed.  We address this in more detail in Appendix B where we derive the perturbation equations by introducing a suitable dimensionless coordinate system that allows us to impose boundary conditions at the correct locations.  In the following subsections (3.1--3.3) we solve the perturbation equations from first to third order.  The reader interested in the results may skip the details and proceed to Section 3.4 for the perturbative analytical solutions in $(b^2,h^2)$ space describing the warped branch.

\newsubsection{First-order perturbations}
\pdfbookmark[2]{First-order perturbations}{First-order perturbations}

The first-order perturbations of \Eqref{eom1} is (see Appendix B)
\begin{align}
 \sbrac{(1-x^2)\frac{\dif^2}{\dif x^2}-qx\frac{\dif}{\dif x}+\rho_0^2\beta}\Phi_1=\rho_0^2F_1\,,\label{gegenbauer}
\end{align}
where we are using the coordinate $x=\sin\ttau$, and
\begin{align}
 \beta&= \frac{2}{q-2}\sbrac{(p-1)(p+q-2)h_0^2+\frac{p(q-1)^2}{p+q-2}b_0^2 - \frac{2p\Lambda}{p+q-2}}\,,\nonumber\\
 F_1  &= (p-1)w_1 + \frac{q-1}{p+q-2}\,.
\end{align}
where we recognize Eq.~\Eqref{gegenbauer} as an inhomogeneous Gegenbauer differential equation with a constant source term (see, e.g., \cite{Gradshteyn:2007,Edelen:1968,Morse:1963}).  Regular solutions exist if $\rho_0$ and $\beta$ satisfy the relation
\begin{align}
 \rho_0^2\beta=l(l+q-1)\,, \hspace{11pt}\mbox{where $l$ is an integer.}
\end{align}
When the above equation is solved for $b_0^2$, and using Eq.~\Eqref{FRbranch1} to eliminate $h_0^2$, we obtain
\begin{align}
 b_0^2=\frac{2\Lambda\sbrac{l(l+q-1)-2q+2}}{(p-1)\sbrac{2q^2-4q+2-l(l+q-1)}}\,,
\end{align}
which is precisely the saturation point of Eq.~\Eqref{instab}, i.e., the points of marginal stability $b^2_\mathrm{m}$.  Therefore we have obtained the first result from our perturbation analysis: \emph{the warped branch, as specified approximately by Eq.~\Eqref{delta_b2}, emanates from the FR branch at $b_0=b_\mathrm{m}$.}
\par
The explicit solution to Eq.~\Eqref{gegenbauer} is given by
\begin{align}
 \Phi_1(x)=k_1C^{(\lambda)}_l(x) + \frac{F_1}{\beta}\,, \label{phi1}
\end{align}
where $\lambda=(q-1)/2$ and $C^{(\lambda)}_l(x)$ is the Gegenbauer polynomial of degree $l$ and order $\lambda$.  $k_1$ is an integration constant determined by the boundary conditions $\Phi(0)=0$.
\par
To complete the first-order analysis, we also seek the first-order solution of \Eqref{eom2}, which will be important in the subsequent sections that follow.  To first order, \Eqref{eom2} is
\begin{align}
 (1-x^2)\frac{\dif^2 a_1}{\dif x^2}+x\frac{\dif a_1}{\dif x} - a_1 = 0\,. \label{eom2-1}
\end{align}
We see that \Eqref{eom2-1} requires $a_1(x)$ to be even about the equator.  This imposes a boundary condition on its first derivative, namely $a_1'(0)=0$.  Together with $a_1(0)=\rho_1$, the solution that satisfies the boundary condition is
\begin{align}
 a_1(x)=\rho_1\brac{\sqrt{1-x^2}+x\arcsin{(x)}}\,. \label{a1}
\end{align}
The explicit value of $\rho_1$ can calculated by evaluating Eq.~\Eqref{constraint} at $a(x)\vert_{x=0}$ up to first order.  Finally, using the above expression of $a_1(x)$, we can obtain the function which we call $X_1$ in Appendix B (Eq.~\Eqref{Xseries}),
\begin{align}
  X_1=-\frac{\rho_1}{\rho_0}\arcsin{(x)}\sqrt{1-x^2}=-\frac{\sqrt{1-x^2}}{\rho_0}\frac{\dif a_1}{\dif x}\,. \label{X1soln}
\end{align}

\newsubsection{Second-order perturbations}
\pdfbookmark[2]{Second-order perturbations}{Second-order perturbations}

The value of $w_1$ is determined by the requirement that the solution $\Phi_2$ of the second order perturbation must be regular at the poles.  To obtain this solution, we expand \Eqref{eom1} to second order by substituting Eq.~\Eqref{phinewseries} into \Eqref{eom1} and collecting the second-order terms.
\begin{align}
  &\sbrac{(1-x^2)\frac{\dif^2}{\dif x^2}-qx\frac{\dif}{\dif x}+\rho_0^2\beta}\brac{\Phi_2+X_1\frac{\dif\Phi_1}{\dif x}} \nonumber\\
  &\hspace{1.5cm}=\frac{2\rho_0^2}{(q-2)^2}\sbrac{(p-1)(p+q-2)h^2_\mathrm{m}+\frac{p^2(q-1)^3}{p+q-2}b_\mathrm{m}^2-\frac{2p^2\Lambda}{p+q-2}}\Phi_2^2\nonumber\\
  &\hspace{2cm}-\frac{2\rho_0^2}{q-2}\sbrac{\frac{p(q-1)^2}{p+q-2}+(p-1)(p+q-2)w_1}\Phi_1+(p-1)\rho_0^2w_2 \nonumber\\
  &\hspace{2cm}-(q-1)\frac{\rho_1}{\rho_0}\brac{x+\frac{\arcsin (x)}{\sqrt{1-x^2}}}\frac{\dif\Phi_1}{\dif x}\,. \label{eom1-2raw}
\end{align}
Using the results obtained in the previous section for $\Phi_1$ and $X_1$ (Eq.~\Eqref{phi1} and \Eqref{X1soln} respectively), the second term in the parentheses on the LHS can be differentiated explicitly, and the resulting expression will cancel some terms on the RHS to give the following resulting equation
\begin{align}
  \sbrac{(1-x^2)\frac{\dif^2}{\dif x^2}-qx\frac{\dif}{\dif x}+\rho_0^2\beta}\Phi_2=\rho_0^2F_2(x)+\rho_0^2\brac{2\frac{\rho_1}{\rho_0}F_1+(p-1)w_2}\,, \label{eom1-2}
\end{align}
where
\begin{align}
  F_2(x)\,=\,&\frac{2}{(q-2)^2}\sbrac{(p-1)(p+q-2)h_\mathrm{m}^2+\frac{p^2(q-1)^3}{p+q-2}b_\mathrm{m}^2-\frac{2p^2\Lambda}{p+q-2}}\Phi_1^2\nonumber\\
         &-\frac{2}{q-2}\sbrac{(p-1)(p+q-2)w_1+\frac{p(q-1)^2}{p+q-2}}\Phi_1-\frac{2\rho_1}{\rho_0}\Phi_1\,.
\end{align}
This is again an inhomogeneous Gegenbauer equation, where the source term $F_2(x)$ is now a function of $x$.  This equation can be solved using the Green's function 
\begin{align}
 G_l(x,x')=\sum_j\frac{j!(\lambda+j)\Gamma(\lambda)^2}{\pi 2^{1-2\lambda}\Gamma(j+2\lambda)}\frac{C^{(\lambda)}_j(x)C^{(\lambda)}_j(x')(1-x'^2)^{\lambda-1/2}}{l(l+q-1)-j(j+q-1)}\,, \label{green}
\end{align}
which allows us to construct a complete solution to Eq.~\Eqref{eom1-2} as follows:
\begin{align}
  \Phi_2(x)= k_2C^{(\lambda)}_{l}(x)+\frac{2\rho_1}{\rho_0}\frac{F_1}{\beta}+\frac{(p-1)w_2}{\beta}+\rho_0^2\int^{1}_{-1}\dif x'\;G_l(x,x')F_2(x')\,.
\end{align}
Regular solutions exist only if $F_2(x)$ satisfies (see, e.g., \cite{Riley:2002} or any standard mathematics textbook)
\begin{align}
 \int^{1}_{-1}\dif x\;F_2(x)C^{(\lambda)}_l(x)\brac{1-x^2}^{\lambda-1/2}=0\,. \label{F2green} 
\end{align}
The remaining unspecified parameter in $F_2(x)$ is $w_1$, which we may adjust to satisfy Eq.~\Eqref{F2green}.  For $l=2N$, $N=1,2,\ldots$ we calculate $w_1$ explicitly by direct integration of Eq.~\Eqref{F2green} using the various properties of Gegenbauer polynomials \cite{Gradshteyn:2007}.  This gives
\begin{align}
 w_1=\frac{h^2_\mathrm{m}+(q-1)Q}{pb^2_\mathrm{m}-(p-1)(p+q-2)Q}\,,  \label{w1levenformula}
\end{align}
where
\begin{align}
  Q\eq&\frac{(-1)^N}{4}\frac{\Gamma(\lambda+N)^2\Gamma(2\lambda+3N)\Gamma(2N+1)(\lambda+2N)}{\Gamma(N+1)^2\Gamma(\lambda)\Gamma(\lambda+3N+1)\Gamma(2\lambda+2N)} \nonumber\\
     &\times\frac{\sbrac{b^2_\mathrm{m}+(p-1)h^2_\mathrm{m}}\sbrac{p^2q(q-1)b^2_\mathrm{m}+(p-1)(p+q-2)(2p+q-2)h^2_\mathrm{m}}}{\Lambda (q-2)^2(p-1)^2}\,.
\end{align}
\par
Before concluding this section, we complete the second-order analysis by obtaining the solution of $a_2$.  This solution will be required in the following section.  Equation \Eqref{eom2} to second order is
\begin{align}
 (1-x^2)&\frac{\dif^2 a_2}{\dif x^2}+x\frac{\dif a_2}{\dif x}-a_2\nonumber\\
     &\;\;=\frac{\rho_1^2}{\rho_0}\brac{\frac{(\arcsin{(x)})^2}{\sqrt{1-x^2}}-\sqrt{1-x^2}}-\frac{p(p+q-2)}{(q-2)^2}\rho_0(1-x^2)^{3/2}\brac{\frac{\dif\Phi_1}{\dif x}}^2\,. \label{eom2-2}
\end{align}
Unlike the earlier case of $a_1$, the equation will be different for each particular $l$, because the source terms depend on $\Phi_1$. The second-order equation is solved by
\begin{align}
 a_2(x)=&\;\;c_1x+c_2\sbrac{\sqrt{1-x^2}+\brac{\arcsin{(x)}-\frac{\pi}{2}}x}-\frac{\rho_1}{2\rho_0}\arcsin{(x)}\sqrt{1-x^2}\frac{\dif a_1}{\dif x}\nonumber\\
        &\;\;+\frac{p(p+q-2)}{(q-2)^2}\rho_0\Biggl[\brac{\sqrt{1-x^2}+x\arcsin{(x)}}\int\dif x\;x\brac{\frac{\dif\Phi_1}{\dif x}}^2 \nonumber\\
        &\;\;\hspace{3cm}-x\int\dif x\;\brac{\sqrt{1-x^2}+x\arcsin{(x)}}\brac{\frac{\dif\Phi_1}{\dif x}}^2\Biggr]\,.
\end{align}
Here $c_1$ and $c_2$ are the integration constants to be determined from the boundary conditions.  Similar to the first-order case, Eq.~\Eqref{eom2-2} is symmetric about the equator, which gives the same boundary condition $a_2'(0)=0$.  The second integration constant can be fixed using $a_2(0)=\rho_2$.
\par
Finally, the function $X_2$ will accordingly be the second-order term of the expansion of Eq.~\Eqref{bigX}.  The specific form will depend on the particular branch, because it depends on $a_1$, $a_2$, and $\Phi_1$; therefore will be too cumbersome to be shown here.  Nevertheless it is a straightforward calculation once the expressions for $a_1$, $a_2$, and $\Phi_1$ are at hand.

\newsubsection{Third-order perturbations}
\pdfbookmark[2]{Third-order perturbations}{Third-order perturbations}

It is evident by now that to analytically obtain the coefficient $w_i$, it is necessary to solve the perturbation equations up to $(i+1)$-th order.  To obtain the third-order equations, we proceed in a manner completely analogous to the procedure in the second-order perturbations.  We substitute \Eqref{phinewseries} into \Eqref{eom1} and collect the terms which are cubic in $\delta b^2$ to produce the third-order analogue of Eq.~\Eqref{eom1-2raw}.  This equation will depend on the functions $\Phi_1$, $\Phi_2$, $a_1$, $a_2$, $X_1$, and $X_2$, which have been obtained in the previous sections.  Substituting these results into the third-order equations gives the corresponding analogue of Eq.~\Eqref{eom1-2}, after some simplification,
\begin{align}
 (1-x^2)\frac{\dif^2\Phi_3}{\dif x^2}-qx\frac{\dif\Phi_3}{\dif x}+\rho_0^2\beta\Phi_3=\rho_0^2F_3(x)-\rho_0^2\brac{\frac{2\rho_0\rho_2+\rho_1^2}{\rho_0}F_1+\frac{2\rho_1}{\rho_0}(p-1)w_2}\,,
\end{align}
where
\begin{align}
 F_3(x)=&\;\;\frac{1}{6}B_1\Phi_1^3+\brac{\frac{2\rho_1}{\rho_0}\beta-B_2+B_3}\brac{\Phi_2+\frac{\rho_1}{\rho_0}(\arcsin{(x)})\sqrt{1-x^2}\frac{\dif\Phi_1}{\dif x}} \nonumber\\ 
         &\;\;+\brac{\frac{2f(x)\sqrt{1-x^2}}{\rho_0x^2}}\brac{F_1-\beta\Phi_1}+\frac{p(p+q-2)}{(q-2)^2}\frac{(1-x^2)^2}{\rho_0^2x}\brac{\frac{\dif\Phi_1}{\dif x}}^3\nonumber\\
         &\;\;+\brac{\frac{2(p-1)(p+q-2)}{q-2}w_2+\frac{2\rho_1}{\rho_0}B_3+\frac{2\rho_0\rho_2+\rho_1^2}{\rho_0^2}\beta}\Phi_1\nonumber\\
         &\;\;-\brac{\half B_4+\frac{\rho_1}{\rho_0}B_2}\Phi_1^2  -\frac{2f(x)\sqrt{1-x^2}}{\rho_0^3x^3}\frac{\dif\Phi_1}{\dif x} \;,
\end{align}
with the following constants abbreviated as
\begin{align}
 B_1=&\;\;(p-1)\brac{\frac{2(p+q-2)}{q-2}}^3\hm+\frac{q-1}{p+q-2}\brac{\frac{2p(q-1)}{q-2}}^3\bm-\frac{2\Lambda}{p+q-2}\brac{\frac{2p}{q-2}}^3\,,\nonumber\\
 B_2=&\;\;(p-1)\brac{\frac{2(p+q-2)}{q-2}}^2\hm+\frac{q-1}{p+q-2}\brac{\frac{2p(q-1)}{q-2}}^2\bm-\frac{2\Lambda}{p+q-2}\brac{\frac{2p}{q-2}}^2\,,\nonumber\\
 B_3=&\;\;(p-1)\brac{\frac{2(p+q-2)}{q-2}}w_1+\frac{q-1}{p+q-2}\brac{\frac{2p(q-1)}{q-2}}\,,\nonumber\\
 B_4=&\;\;(p-1)\brac{\frac{2(p+q-2)}{q-2}}^2w_1+\frac{q-1}{p+q-2}\brac{\frac{2p(q-1)}{q-2}}^2\,,
\end{align}
and the function $f(x)$ is given by
\begin{align}
 f(x)=\frac{p(p+q-2)}{(q-2)^2}\rho_0\sqrt{1-x^2}\int\dif x\;x\brac{\frac{\dif \Phi_1}{\dif x}}^2\,.
\end{align}
The particular solution $\Phi_3$ can be evaluated analytically by a Green's function integral similar to Eq.~\Eqref{F2green}, where for regular solutions, we require
\begin{align}
 \int^{1}_{-1}\dif x\;F_{3}(x)C^{(\lambda)}_l(x)\brac{1-x^2}^{\lambda-1/2}=0\,, \label{F3green} 
\end{align}
giving the value of $w_2$.

\newsubsection{Results}
\pdfbookmark[2]{Results}{Results}

Summarizing the results on the previous sections, we find that the phase space of the warped branches can be approximately described by
\begin{align}
  h^2=h^2_\mathrm{m}+w_1\delta b^2+w_2\brac{\delta b^2}^2+\mathcal{O}\sbrac{\brac{\delta b^2}^3}\,,
\end{align}
where the values $w_1$ and $w_2$ are derived in the previous sections.  We list the results in Tables \ref{w1} and \ref{w2} for $p=4$, and the cases $q=4$, $5$, $6$ and $7$.
\begin{table}
\begin{center}
\renewcommand{\arraystretch}{1.5}
 \begin{tabular}{ccccc}
	\hline
       $q$:  & $4$             & $5$                            & $6$                           & $7$  \\ \hline
     $l=2$:  & $-\frac{7}{54}$& $-\frac{431}{3444} $          & $-\frac{199}{1668}$          & $-\frac{8}{71} $   \\
     $l=4$:  & $ $             & $ $                            & $-\frac{2293}{10056}$        & $-\frac{646}{2547}$ \\  \hline
 \end{tabular}
 \caption{Exact values of $w_1$ for $p=4$ and various $q$, for even $l$ branches}
 \label{w1}
\end{center}
\end{table}

\begin{table}
\begin{center}
\renewcommand{\arraystretch}{1.5}
 \begin{tabular}{ccccc}
	\hline
       $q$:  & $4$                  & $5$                      & $6$                                 & $7$  \\ \hline
     $l=2$:  & $-\frac{722}{216513}$& $-\frac{315975}{9790144}$& $-\frac{91208835}{1117217504}$     & $-\frac{197372}{1073733} $   \\
     $l=4$:  & $ $                  & $ $                      & $-\frac{24624514905}{3041267079296}$& $-\frac{100489280}{1835880147}$ \\  \hline
 \end{tabular}
 \caption{Exact values of $w_2$ for $p=4$ and various $q$, for even $l$ branches}
 \label{w2}
\end{center}
\end{table}
Note that the values of $w_2$ are nonzero, which shows that the warped branch is not purely linear in the $(b^2,h^2)$ phase space.  For example, in the case studied by Kinoshita ($p=q=4$, $l=2$), the warped branch is (in units where $\Lambda=1$)
\ba
  h^2=\frac{1}{18} - \frac{7}{54}\delta b^2 -\frac{722}{216513}\brac{\delta b^2}^2+\mathcal{O}\sbrac{\brac{\delta b^2}^3}.
\ea
For this particular case, we have $h^2_\mathrm{m}=1/18$ in the first term which is in expected agreement with \cite{Bousso:2002fi,Martin:2004wp}.  The next two terms are with $w_1=-7/54$ and $w_2=-722/216513$  which are precisely obtained from Eqs.~\Eqref{w1levenformula} and \Eqref{F3green} respectively.  It has been checked that the results of Tables \ref{w1} and \ref{w2} agree with the numerical results within numerical accuracy.

\newsection{Odd $l$ cases}

For the odd $l$ cases, if we apply the procedure of the previous section, it may appear that the perturbative solutions to Eqs.~\Eqref{eom1} and \Eqref{eom2} could not be determined  uniquely.  The reason for this can be seen from Eq.~\Eqref{phi1}.  If $l$ is odd, then $C^{(\lambda)}_l(x)$ is anti-symmetric, thus vanishing identically at $x=0$.  This implies that at the equator, $\Phi_1(x=0)=F_1/\beta\neq 0$, which is inconsistent with the choice in Eq.~\Eqref{scheme1}.
\par
To overcome this problem, we exploit the freedom given by Eq.~\Eqref{invariance} and shift the zero point of $\Phi$ by choosing $\Phi(x=-1)=0$ (this is equivalent to $\Phi(\theta_-)=0$, i.e., $\Phi$ is zero at the south pole) and perform the perturbation analysis in this scheme.  The results can be consistently transformed back via the inverse of Eq.~\Eqref{invariance} to obtain the approximate solutions for the warped branches.  Such transformations must be handled with care; the transformation \Eqref{invariance} involves a rescaling of $b^2$.  Since $b^2$ is our perturbation parameter, different perturbation orders will be mixed up by the transformation \Eqref{invariance}.
\par
Indeed, we find that information from the second-order perturbation results under the scheme $\Phi(\theta_-)=0$ is transformed into first-order under the scheme $\Phi(\theta_*)=0$. Therefore perturbations up to third-order, if converted back to the choice $\Phi_1(x=0)=F_1/\beta\neq 0$ only gives values up to $w_1$.  To obtain $w_2$ for odd $l$ cases one may need to go beyond third-order perturbations, which is presently beyond the scope of our analysis. The numerical values for $w_1$ in the odd $l$ branches are given in Table \ref{w1odd}.  For example, in the case $p=4$, $q=5$ the phase diagram for the warped branch is given by (in units where $\Lambda=1$)
\begin{align}
 h^2=\frac{22}{315}-\frac{174231551}{577251843}\delta b^2+\mathcal{O}\sbrac{\brac{\delta b^2}^2}\,.
\end{align}
Here we have $h^2_\mathrm{m}=22/315$, in agreement with \cite{Bousso:2002fi,Martin:2004wp}, and the next term is with $w_1=-174231551/577251843$.
\begin{table}
\begin{center}
\renewcommand{\arraystretch}{1.5}
 \begin{tabular}{ccccc}
	\hline
       $q$:  & $5$                            & $6$                           & $7$  \\ \hline
     $l=3$:  & $-\frac{174231551}{577251843}$& $-\frac{56669459}{166268616}$& $-\frac{5316380}{15099489}$    \\
     $l=5$:  & $ $                            & $ $                           & $-\frac{5029755010667}{24287636988414}$ \\ \hline
 \end{tabular}
 \caption{Exact values of $w_1$ for $p=4$ and various $q$, for odd $l$ branches}
 \label{w1odd}
\end{center}

\end{table}

\newsection{Thermodynamics}
\pdfbookmark[2]{Thermodynamics}{Thermodynamics}

As an application and consistency check of our perturbation results, we turn to the study of de Sitter thermodyamics. Below we will use our perturbation equations to derive the first law of thermodynamics, which was first derived by \cite{Kinoshita:2009hh} using a different method.  Clearly, the following analysis will only apply to solutions containing a horizon, i.e., $h^2>0$. 
\par
In \cite{Kinoshita:2009hh}, Kinoshita and Mukohyama defined the entropy $\mathcal{S}$ and total flux $\mathcal{F}$ as the basic thermodynamic variables:
\begin{align}
  \mathcal{S}&=\frac{\Omega_{(p-2)}\Omega_{(q-1)}}{4h^{p-2}}\int\dif\theta\;a^{q-1}\expo{-\frac{2(p+q-2)}{q-2}\Phi}\,, \label{entropy}\\
  \mathcal{F}&=\oint F_{(q)}=b\Omega_{(q-1)}\int\dif\theta\;a^{q-1}\expo{-\frac{2p(q-1)}{q-2}\Phi}\,.  \label{flux}
\end{align}
By a scaling argument and variation of the action \Eqref{action} (see Ref.~\cite{Kinoshita:2009hh} for details), the following first law of thermodynamics was derived:
\begin{align}
  \dif\mathcal{S}=-\frac{\Omega_{(p-2)}b}{4(p-1)h^p}\dif\mathcal{F}\,. \label{firstlaw}
\end{align}
We can show that Eq.~\Eqref{firstlaw} can also be derived from the perturbation analysis.  We consider the changes in $\mathcal{S}$ and $\mathcal{F}$ as the warped solution moves away from the marginal stability points. Hence the parameters changes according to Eq.~\Eqref{delta_b2}, which we rewrite here for convenience:
\begin{align}
 b^2&=b^2_\mathrm{m}+\delta b^2,\hspace{0.5cm} h^2=h_\mathrm{m}^2+w_1\delta b^2+\ldots
\end{align}
Here the subscripts in $\mathcal{S}_\mathrm{w}$ and $\mathcal{F}_\mathrm{w}$ indicates that we are referring to the entropy and flux for the warped branch.  Recall that when $\delta b^2=0$, the warped and FR branches coincide, so at $\delta b^2=0$, we have $\mathcal{S}_\mathrm{w}=\mathcal{S}_\mathrm{FR}\equiv\mathcal{S}_\mathrm{m}$ and $\mathcal{F}_\mathrm{w}=\mathcal{F}_\mathrm{FR}\equiv\mathcal{F}_\mathrm{m}$.
For $\delta b^2\neq 0$, small increments to $\delta b^2$ induces corresponding changes to the entropy and flux,
\begin{align}
  \mathcal{S}_\mathrm{w}=\mathcal{S}_\mathrm{m}+\delta\mathcal{S}_\mathrm{w}\,, \hspace{0.5cm}\mathcal{F}_\mathrm{w}=\mathcal{F}_\mathrm{m}+\delta\mathcal{F}_\mathrm{w}\,. \label{dSdF}
\end{align}
To calculate $\delta\mathcal{S}_\mathrm{w}$ and $\delta\mathcal{F}_\mathrm{w}$, we expand \Eqref{entropy} and \Eqref{flux} to first order and make use of the results  \Eqref{phi1} and \Eqref{a1}.  Evaluating the resulting integrals, we find, to first order
\begin{align}
  \delta\mathcal{S}_\mathrm{w}=\;&\frac{\Omega_{(p-2)}\Omega_{(q-1)}}{4h_\mathrm{m}^{p-2}}\frac{\sqrt{\pi}\Gamma\brac{\frac{q}{2}}}{\Gamma\brac{\frac{q+1}{2}}}\rho_0^q\biggl\{(q-1)\frac{\rho_0}{\rho_1}-\frac{2(p+q-2)F_1}{\beta}-\frac{(p-2)w_1}{2h^2_\mathrm{m}} \nonumber\\
                  &+(q-1)\frac{\rho_1}{\rho_0}\frac{\sqrt{\pi}\Gamma(q-1)}{2^{q-1}\Gamma\brac{\frac{q}{2}}\Gamma\brac{\frac{q-1}{2}}}\sbrac{\psi\brac{\frac{q+1}{2}}-\psi\brac{\frac{q-1}{2}}}\biggr\} \delta b^2\,, \label{dS}
\end{align}
where $\psi(x)$ is the Euler psi function defined by $\psi(x)=\frac{\dif}{\dif x}\ln\Gamma(x)$ \cite{Gradshteyn:2007}.  Similarly, the first order change in flux is
\begin{align}
  \delta\mathcal{F}_\mathrm{w}=\;&b_\mathrm{m}\Omega_{(q-1)}\frac{\sqrt{\pi}\Gamma\brac{\frac{q}{2}}}{\Gamma\brac{\frac{q+1}{2}}}\rho_0^q\biggl\{\frac{1}{2b_\mathrm{m}^2}+(q-1)\frac{\rho_0}{\rho_1}-\frac{2p(q-1)F_1}{\beta} \nonumber\\
                  &+\frac{\rho_1}{\rho_0}\frac{\sqrt{\pi}\Gamma\brac{q-1}}{2^{q-1}\Gamma\brac{\frac{q}{2}}\Gamma\brac{\frac{q-1}{2}}}\sbrac{\psi\brac{\frac{q+1}{2}}-\psi\brac{\frac{q-1}{2}}}\biggr\} \delta b^2\,.  \label{dF}
\end{align}
Eliminating $\delta b^2$ (with the aid of Eq.~\Eqref{FRbranch1}) Eqs.~\Eqref{dS} and \Eqref{dF} reduces to \Eqref{firstlaw}, reproducing the first law (up to first order, we have $\delta\mathcal{S}=\dif\mathcal{S}$ and $\delta\mathcal{F}=\dif\mathcal{F}$). This demonstrates the consistency of the perturbative analysis with the first law of thermodynamics.  Due to the fact that both the FR and warped branches obey Eq.~\Eqref{firstlaw}, it follows from Eq.~\Eqref{dSdF} that at first order, the FR and warped solution with the same flux also have the same entropy.  It is reminiscent of the analysis of black brane stability, where for the same mass, the uniform and non-uniform black branes have the same entropy up to first order \cite{Gubser:2002}.
\par
If we extend the comparison by including second-order terms, we find that
\begin{align}
  \mathcal{S}_\mathrm{w}-\mathcal{S}_\mathrm{FR}\arrowvert_{\mathcal{F}_\mathrm{w}=\mathcal{F}_\mathrm{FR}}\sim \mathcal{O}\sbrac{\brac{\delta b^2}^3}\,,
\end{align}
which means the respective entropies at second order are still the same.  When the difference in the above equation is calculated, there is a nonzero term of order $\brac{\delta b^2}^3$, which may possibly indicate that the entropy values for the different branches begin to differ at third order.  However to verify this one has to calculate the expansions of $\mathcal{S}$ and $\mathcal{F}$ up to third order, which requires the exact solution for $a_3(x)$, which we do not have.  
\par
We may compare the above analytical calculations with the numerical results of \cite{Kinoshita:2009hh}.  The entropy difference $\mathcal{S}_\mathrm{w}-\mathcal{S}_\mathrm{FR}$ for fully non-perturbative (numerical) solutions of Eqs.~\Eqref{eom1} and \Eqref{eom2} was calculated in \cite{Kinoshita:2009hh} and shown to be consistent with the correlated stability conjecture \cite{Gubser:2000mm}, i.e., $\mathcal{S}_\mathrm{w}>\mathcal{S}_\mathrm{FR}$ when the FR branch is dynamically unstable.  An inspection of Figure 3 of Ref.~\cite{Kinoshita:2009hh} shows that $\mathcal{S}_\mathrm{w}-\mathcal{S}_\mathrm{FR}$ as a function $\mathcal{F}$ appears flat for regions close to the marginal stability points.  Our calculations above identify the scope of this region to be at least second order in deviations away from the marginal stability point.

\newsection{Conclusion}

In this work, we have used Kinoshita's ansatz to find numerical solutions of various modes for the cases $q=4$, $5$, $6$ and $7$.  It was shown that the warped branches under this ansatz are consistent with the stability analysis for the FR branch by Martin \cite{Martin:2004wp} and Bousso \cite{Bousso:2002fi}.  In particular, the points of marginal stability gives rise to a new branch emanating from the FR branch.  Although the warped branch in phase space appears to lie on a straight line, perturbation theory shows that this is not true because $w_2$ is indeed small, but still nonzero.  The warped branch has a quadratic dependence and is not purely linear in $b^2$.
\par
Using perturbation theory, expanding Kinoshita's equations of motion about the FR branch reproduces the points of marginal stability $(b^2_\mathrm{m},h^2_\mathrm{m})$.  Also we were able to derive the explicit solutions of $\Phi$ and $a$ up to second order.  Furthermore it was found that the requirement that the solutions of $\Phi_i$ must be regular provides constraints between $h^2$ and $b^2$, which gives the approximate equations which describe the phase diagram for the warped branches in regions close to the FR branch.
\par
As an application to de Sitter thermodynamics, we calculated the thermodynamic variables $\mathcal{S}$ and $\mathcal{F}$ for the warped branches and reproduced the first law of thermodynamics.  Our calculations show that for the same flux $\mathcal{F}$, the two branches have equal entropy up to second order.  However it is already known from the numerical results of \cite{Kinoshita:2009hh} that the warped branch has higher entropy when the FR branch is dynamically unstable.  Therefore we expect that the entropy difference between the two branches should enter at third-order or beyond.
\par
In addition to the above mentioned results, we have also found numerical evidence for the existence of yet another new type of branch which is not yet captured by any previous study.  This branch intersects with the $q=4$, $l=2$ branch, but does not intersect with the FR branch.  It has less symmetry than the $l=2$ branch (it is neither symmetric nor anti-symmetric about the equator) and satisfies the boundary conditions defined above.  The intersection with the warped $q=4$, $l=2$ branch occurs around the region $h^2\sim 0.0199\Lambda$, $b^2\sim 0.0951\Lambda$, which is where the $l=2$ is unstable and hence beyond the perturbation analysis of Ref.~\cite{Kinoshita:2009hh}. Although not much is known about this branch at this point, it indicates the possibility that warped flux compactifications exhibit a rich structure and there may be many families of solutions which might deserve further study.

\vspace{12mm}
\pagebreak[3]
\noindent{\bf Acknowledgements}
\nopagebreak
\medskip
\nopagebreak

The author would like to thank Edward Teo for useful comments and contributions to this work.

\appendix

\newsection{Numerical results for $q=5,6$ and $7$}

\begin{figure}[ht]
\begin{center}
  \includegraphics[angle=-90,scale=0.6]{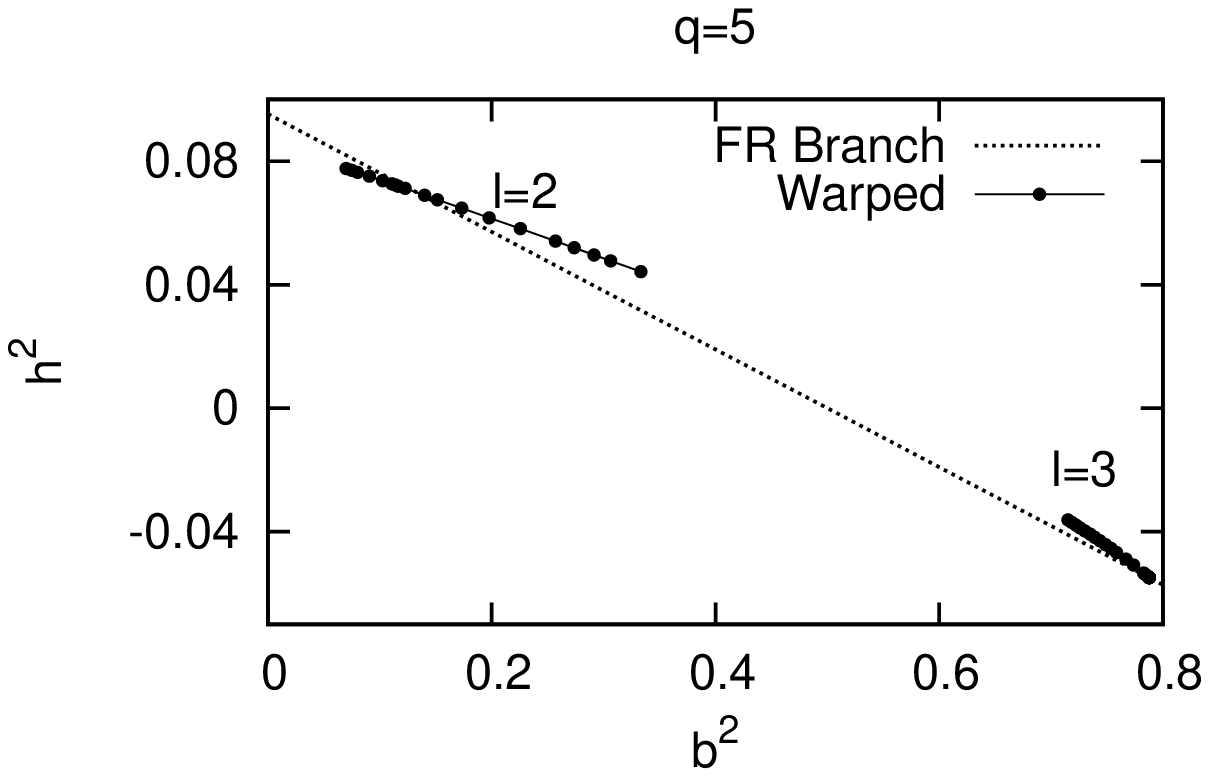}
  \includegraphics[angle=-90,scale=0.6]{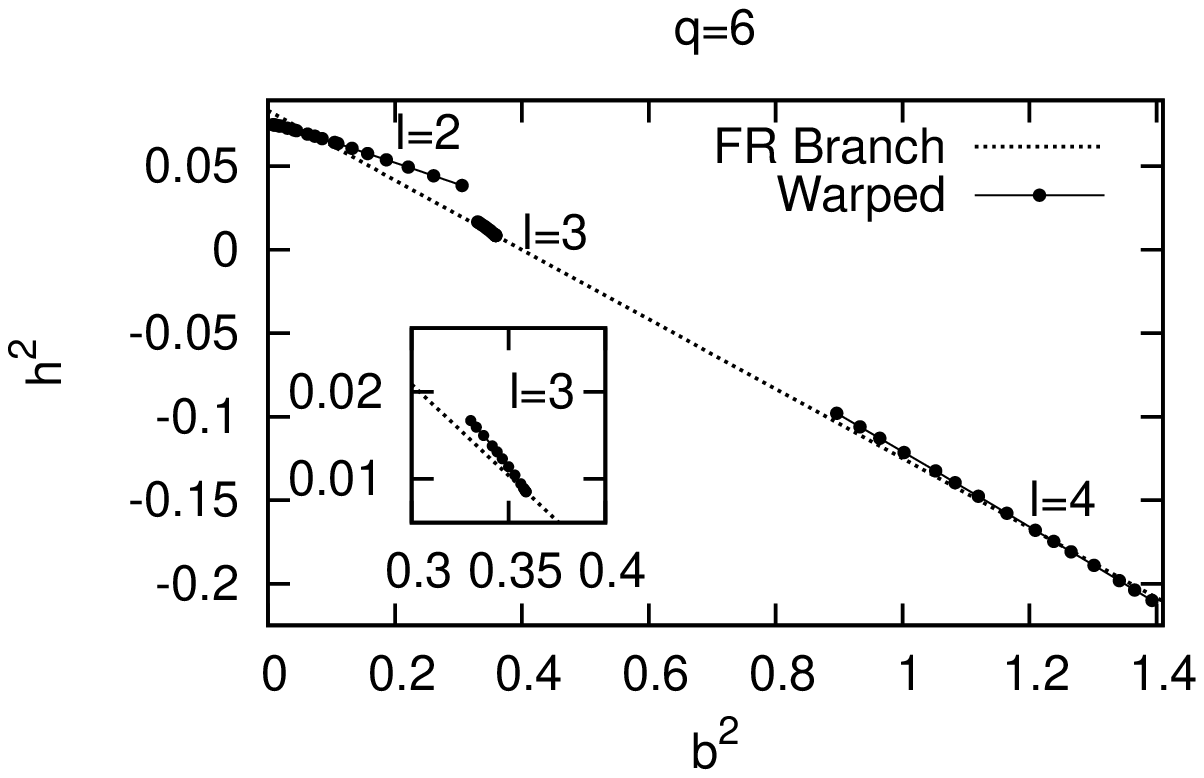}\\
  \includegraphics[angle=-90,scale=0.5]{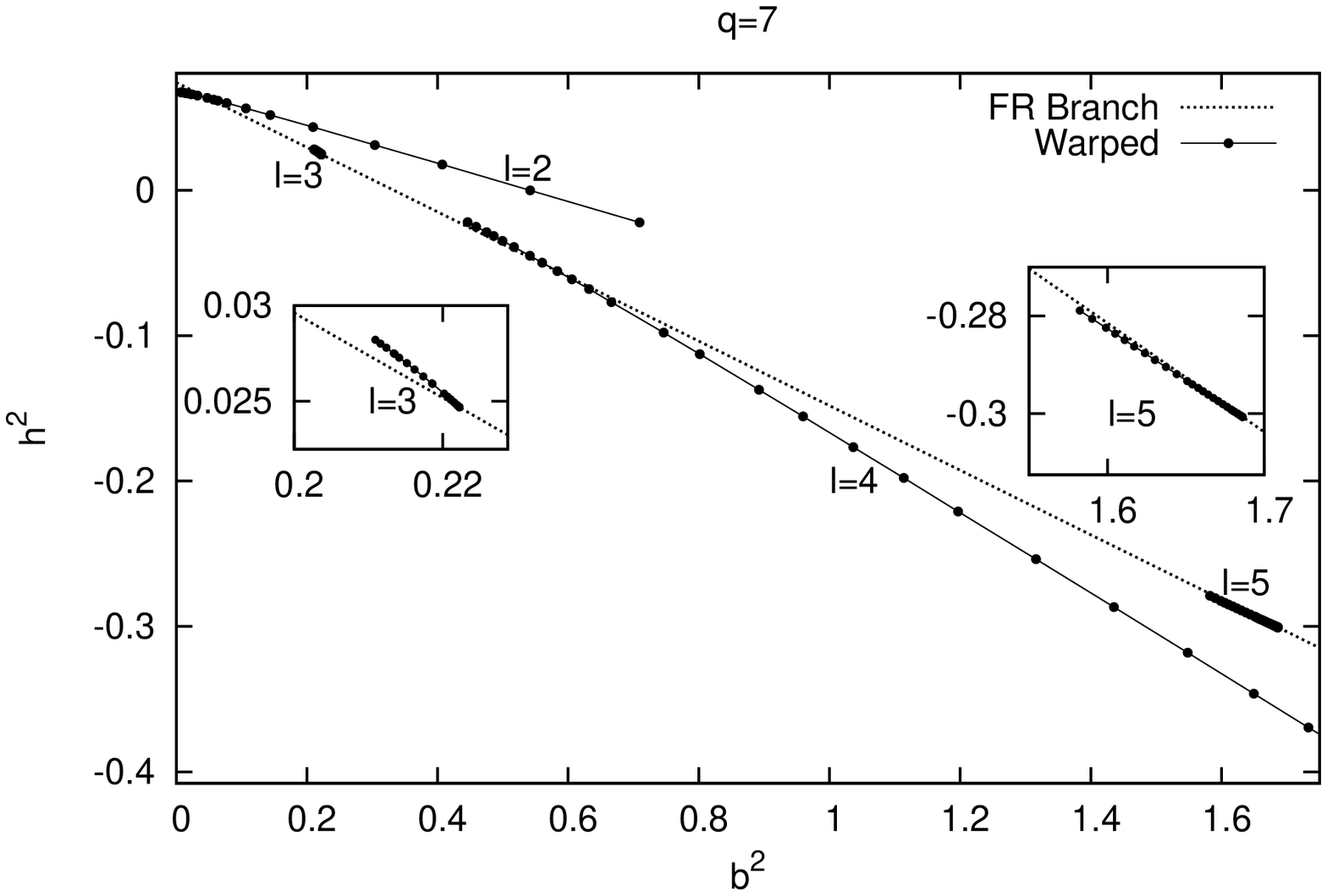}
  \caption{$b^2$ vs $h^2$ phase diagrams for cases $q=5$ (top left), $q=6$ (top right) and $q=7$ (bottom).  The smaller branches for the cases $q=6$ and $q=7$ are magnified in the insets for clarity.}
  \label{q567-branches}
\end{center}
\end{figure}

Eqs.~\Eqref{eom1} and \Eqref{eom2}, in general, can be solved for any choice of $p$ and $q$.  We obtain the solutions for the cases $p=4$, $4\leq q\leq 7$.  For each case, the warped branch intersects with the FR branch at the point of marginal stability $(b^2_\mathrm{m},h^2_\mathrm{m})$.  See Figure \ref{q4branches} for the $q=4$ case \cite{Kinoshita:2007uk} and Figure \ref{q567-branches} for $q=5$, $6$, $7$.
\par
In order to obtain all possible branches, the boundary conditions used to solve the Einstein equations are Eq.~\Eqref{BC}, which are slightly different from the one used by \cite{Kinoshita:2007uk}.\footnote{In \cite{Kinoshita:2007uk}, one of the boundaries is the equator, and the condition imposed is $\Phi'(\theta_*)=0$; which will produce the even $l$ solutions, but not solutions with odd $l$.}  In our case, the Einstein equations are integrated from the north to south pole, and hence, there are no constraints imposed on $\Phi$ at the equator, and thus solutions with $\Phi'(\theta_*)=0$ (even $l$ solutions) and $\Phi'(\theta_*)\neq 0$ (odd $l$ solutions) can be found.  This is a straightforward generalization of the numerics of \cite{Kinoshita:2007uk} which allows us to find the odd $l$ solutions which were previously not considered.
\par
The specific shape of the solutions $\Phi$ will have similar shapes to the particular $l$ modes from which it emanated from the relevant FR instability point, thus we name each branch by their corresponding $l$'s.  As we have seen from Section 3, at regions close to the FR branch, the warped solutions $\Phi$ are approximately Gegenbauer polynomials plus correction terms.  Taking a specific example, in $q=6$, the $l=3$ branch is the one which emanated from the $l=3$ instability $b^2_\mathrm{m}=14\Lambda/39$, $h^2=\Lambda/117$.  Close to the FR branch, $\Phi$ is proportional to $C^{(5/2)}_3(\sin\ttau)$.  In the non-perturbative regime, $\Phi$ will still inherit some features of $C^{(5/2)}_3(\sin\ttau)$, namely it is anti-symmetric about $\theta_*$, and has four turning points.  On the other hand, the solution for $a$ in all cases have the same general appearance as shown in \cite{Kinoshita:2007uk}.
\par

\newsection{Establishing the perturbation equations and boundary conditions}

To derive the perturbation equations of \Eqref{eom1} and \Eqref{eom2}, we need to establish a suitable coordinate system to remove the ambiguity of the boundary locations.  Our starting point is to first consider the FR branch described by Eq.~\Eqref{FRsoln}.  It is intuitively clear that $\theta/\rho_0$ is a dimensionless angular parameter that labels the points along one direction of the $q$-sphere with radius $\rho_0$.  This also means that $\theta$ has dimension of length and takes values $-\pi\rho_0/2\leq\theta\leq\pi\rho_0/2$.
\par
Since we are looking for a warped solution by perturbing the FR branch, the length scale associated with $\theta$ will change when the FR solution is deformed.  This introduces an ambiguity in the endpoints of $\theta$ (which are the location of the poles).  This is problematic because there is now an ambiguity in the location for which to impose the boundary conditions.   To approach this perturbation systematically, it is convenient to define
\begin{align}
  X=\sqrt{1-\frac{a(\theta)^2}{a(\theta_*)^2}}\,.\label{bigX}
\end{align}
Since for both the warped and FR branches, boundary conditions require $a(\theta)$ to vanish at the poles.  So the poles can be specified precisely as $a(X)\arrowvert_{X=\pm 1}=0$.  For the case of the FR branch, $a_\mathrm{FR}(\theta)=\rho_0\cos\ttau$, we simply have $X=\sin\ttau$.  The coordinate $X$ for the case $a=a_\mathrm{warped}$ is expected to be the corresponding generalization to include the warped branch.  Recall that when perturbing the FR branch to find approximate warped solutions, $a(\theta)$ is written as a series in Eq.~\Eqref{alphaseries}.  Inserting Eq.~\Eqref{alphaseries} into \Eqref{bigX} and expanding will give $X$ as a series which we denote by
\begin{align}
  X=\rho_0\sin\ttau+X_1\delta b^2+X_2\brac{\delta b^2}^2+\ldots\,, \label{Xseries}
\end{align}
where $X_1, X_2, \ldots$ are the coefficients of the respective orders of the expansion.  We find it convenient to use $x=\sin\ttau$ as our coordinate in the differential equations.  With the inverse relation $\theta=\rho_0\arcsin (x)$, the $X$ defined in Eq.~\Eqref{bigX}, and its expansion \Eqref{Xseries} can be viewed as functions of $x$.  Therefore the perturbation series of the warp factors can be calculated with $x$ as a variable ($-1\leq x\leq 1$), with corrections due to the shifting of the poles correctly taken into account when $X$ is evaluated up to appropriate order.
\par
Using our new coordinate system, we can now write the series in Eq.~\Eqref{phiseries} as follows,
\begin{align}
  \Phi\brac{X}=\;&\Phi\brac{x+X_1\delta b^2+X_2\brac{\delta b^2}^2+\ldots}\nonumber\\
              =\;&\Phi_1(x)\delta b^2+\brac{\Phi_2(x)+X_1\frac{\dif\Phi_1(x)}{\dif x}}\brac{\delta b^2}^2\nonumber\\
&\hspace{0.5cm}+\brac{\Phi_3(x)+X_1\frac{\dif\Phi_2(x)}{\dif x}+X_2\frac{\dif\Phi_1(x)}{\dif x}+\half X_1^2\frac{\dif^2\Phi_1(x)}{\dif x^2}}\brac{\delta b^2}^3+\mathcal{O}\sbrac{\brac{\delta b^2}^4},\nonumber\\
         \mbox{} \label{phinewseries}
\end{align}
where the second equality follows from Taylor expansion up to third order in $\delta b^2$.  Then, to derive the perturbation equations, we insert \Eqref{phinewseries} into Eq.~\Eqref{eom1} and expand to appropriate order.  The procedure is completely analogous for the perturbations of $a$.

\bigskip\bigskip

{\renewcommand{\Large}{\normalsize}
}

\end{document}